\documentclass[11pt]{article}

\RequirePackage[utf8]{inputenc}
\RequirePackage[T1]{fontenc}

\usepackage{amsmath,amssymb,amsthm,bm,mathtools}
\usepackage{booktabs,tabularx,array,multirow}
\usepackage[table]{xcolor}
\usepackage[protrusion=true,expansion=false]{microtype}
\usepackage{enumitem}
\usepackage{graphicx}
\usepackage{tikz}
\usetikzlibrary{arrows.meta,positioning,fit,calc}
\usepackage{caption}
\usepackage{subcaption}
\usepackage{float}
\usepackage{needspace}
\usepackage{csquotes}
\usepackage{url}
\usepackage[numbers,sort&compress]{natbib}
\usepackage{geometry}
\usepackage{titlesec}
\usepackage{fancyhdr}
\usepackage{lastpage}
\usepackage{hyperref}
\usepackage[nameinlink,capitalise,noabbrev]{cleveref}

\definecolor{deepblue}{HTML}{17365D}
\definecolor{teal}{HTML}{167D8D}
\definecolor{softblue}{HTML}{EAF2F8}
\definecolor{softgray}{HTML}{F3F5F7}
\definecolor{warmgray}{HTML}{555555}

\hypersetup{
  colorlinks=true,
  linkcolor=deepblue,
  citecolor=teal,
  urlcolor=deepblue,
  pdftitle={Quantum-Like Spatial Decision Dynamics: A Falsifiable Model of Cue-Order Effects in Immersive Navigation},
  pdfauthor={Aryabrata Basu},
  pdfkeywords={quantum cognition, spatial decision making, virtual reality, cue-order effects, model comparison, human-quantum interaction}
}

\titleformat{\section}{\Large\bfseries\color{deepblue}}{\thesection}{0.65em}{}
\titleformat{\subsection}{\large\bfseries\color{deepblue}}{\thesubsection}{0.65em}{}
\titleformat{\subsubsection}{\normalsize\bfseries\color{warmgray}}{\thesubsubsection}{0.65em}{}
\setlist{nosep,leftmargin=1.35em}
\newcolumntype{Y}{>{\raggedright\arraybackslash}X}

\newtheorem{proposition}{Proposition}
\newtheorem{definition}{Definition}
\newcommand{\Tr}{\operatorname{Tr}}
\newcommand{\Prob}{\mathbb{P}}
\newcommand{\Hilb}{\mathcal{H}}
\newcommand{\Context}{\mathcal{C}}
\newcommand{\Instrument}{\mathcal{I}}

\newcommand{\ket}[1]{\lvert #1 \rangle}
\newcommand{\bra}[1]{\langle #1 \rvert}

\newcommand{\projector}[1]{\Pi_{#1}}

\title{\textbf{Quantum-Like Spatial Decision Dynamics:}\\
\Large A Falsifiable Model of Cue-Order Effects in Immersive Navigation\\[0.35em]
\large with Implications for Human--Quantum Computer Interaction}
\author{Aryabrata Basu\\
\small School of Computer and Information Sciences\\
\small University of Arkansas at Little Rock\\
\small \href{mailto:abasu@ualr.edu}{abasu@ualr.edu}}
\date{October 2026\\[0.4em]\small Conceptual Framework, Simulation Study, and Proposed Experimental Design}

\begin{document}
\maketitle

\begin{abstract}
The sequence in which a person encounters spatial evidence can change a later choice, yet an order effect alone does not identify a quantum-like cognitive structure. We introduce \emph{Quantum-Like Spatial Decision Dynamics} (QSDD), a falsifiable state-space account of embodied decision making in which cue exposures are completely positive trace-preserving maps, intermediate judgments are quantum instruments, and route commitment is an operationally defined measurement. We distinguish this formal claim from any assertion that cognition is microscopically quantum. We then specify a minimal binary model with a fixed decision basis, noncommuting cue rotations, a fixed symmetry-breaking initial azimuth, dephasing, and lapse; its five fitted parameters must generalize across environments rather than being refit to individual conditions. A reproducible simulation study evaluates recovery and model discrimination under both QSDD and seven-parameter classical logistic ground truths. Across 40 replicates per design cell, QSDD was preferred on held-out environments in 85\% of QSDD-generated datasets at 240 independent observations per environment-order cell and 95\% at 480, but in 0\% of classically generated datasets at every tested sample size. The simulation also exposes weak identification of the lapse parameter and is explicitly a design analysis, not human evidence. We provide a preregistrable virtual-reality experiment, telemetry schema, classical comparison set, and failure criteria. Finally, we show how the same process-measurement logic can be transferred to human inspection and debugging of quantum circuits. QSDD is therefore offered as a constrained model to be defeated or supported by data, and as a methodologically continuous route from immersive interaction research to human--quantum computer interaction.
\end{abstract}


\vspace{0.4em}
\begin{center}
\fcolorbox{deepblue}{softblue}{\parbox{0.92\linewidth}{\textbf{Evidence status.} The numerical results in this manuscript are generated by a fully scripted simulation with known ground truths; no human-subjects data are reported. ``Quantum-like'' denotes Hilbert-space probability and measurement-inspired state update. It does not assert microscopic quantum coherence in neural tissue, quantum consciousness, or a contribution to quantum hardware.}}
\end{center}

\section{Introduction}

Spatial decisions are rarely made from a neutral vantage point. A navigator turns the head, approaches a landmark, opens a map, hears a directional cue, hesitates, backtracks, and eventually crosses a boundary that expresses a choice. Each act changes what information is available and may change the state from which the next judgment is made. The order of observations is therefore not merely a record of a pre-existing decision. It can be constitutive of the decision.

Immersive environments offer unusual experimental leverage over this process. They preserve continuous, embodied interaction while allowing precise control of the scene, cue timing, viewpoint, locomotion, and consequences. They also permit dense measurement of gaze, head and hand motion, path geometry, pauses, selections, and revisits. Virtual reality (VR) has accordingly become an important instrument for spatial-cognition research, although transfer from virtual to physical navigation must be evaluated rather than presumed \citep{hejtmanek2020transfer,draschkow2022frames,yan2026divergent}.

Most computational accounts of spatial choice remain classical. Bayesian models represent belief revision, Markov and partially observable Markov decision processes represent state transition and action, and sequential-sampling models represent noisy evidence accumulation toward a boundary. These are powerful baselines. Yet some human judgments display context dependence, question-order effects, and departures from simple total-probability decompositions for which quantum probability has provided compact formal accounts \citep{pothos2009quantum,busemeyer2011conjunction,pothos2013direction,wang2014order,yearsley2016zeno,pothos2022review}. Quantum cognition employs the mathematics of quantum theory as a probability calculus for cognition; it does not, by itself, claim that cognitive activity is implemented by a physically quantum brain \citep{busemeyer2012quantum,yearsley2016tutorial,khrennikov2015modeling}.

An order effect, however, is not uniquely quantum. A classical model with memory, recency, nonlinear state transition, or condition-specific parameters can also produce sequence sensitivity. The scientific burden is therefore stronger: QSDD must impose a reusable geometry, make cross-condition predictions with frozen operators, survive parameter-recovery tests, and outperform strong classical alternatives on held-out participants and environments. Merely fitting an interference term after observing the data is not evidence.

This paper addresses that burden in three steps. First, it develops a consistent operator account in which cue exposures are channels and registered judgments are instruments. Second, it reduces that account to a five-parameter binary model that can be estimated and falsified. Third, it reports a two-ground-truth simulation showing when model comparison can and cannot distinguish that model from a more flexible classical alternative. The resulting protocol is then carried into an immersive navigation experiment.

The work also has a domain-specific transfer target. Quantum computing increasingly requires interfaces through which learners and practitioners compose, inspect, compare, and debug quantum programs \citep{ashktorab2019thinking,kim2025toward,bano2026human}. QTerra treats a variational quantum circuit as an immersive work object and records how users inspect, revisit, manipulate, and reason through its representations \citep{basu2026qterra}. QSDD can therefore be tested first in controlled spatial decisions and later in circuit-inspection tasks without claiming a new quantum algorithm or a microscopic theory of mind.

\begin{figure}[htbp]
\centering
\begin{tikzpicture}[
  node distance=0.7cm,
  bridge/.style={draw=deepblue,rounded corners=2pt,fill=softblue,align=center,minimum height=1.0cm,text width=2.75cm},
  arr/.style={-{Latex[length=2.1mm]},thick,draw=deepblue}
]
\node[bridge] (formal) {Constrained QSDD\\fixed state, channels, measurement};
\node[bridge,right=of formal] (simulation) {Design analysis\\recovery and two-ground-truth discrimination};
\node[bridge,right=of simulation] (experiment) {Immersive test\\proposed cue-order experiment};
\node[bridge,right=of experiment] (transfer) {Domain transfer\\quantum-circuit inspection and debugging};
\draw[arr] (formal)--(simulation);
\draw[arr] (simulation)--(experiment);
\draw[arr] (experiment)--(transfer);
\end{tikzpicture}
\caption{Evidentiary sequence. Formal specification and simulation precede human testing; quantum-circuit tasks are a transfer domain rather than evidence for the navigation model.}
\label{fig:evidence}
\end{figure}
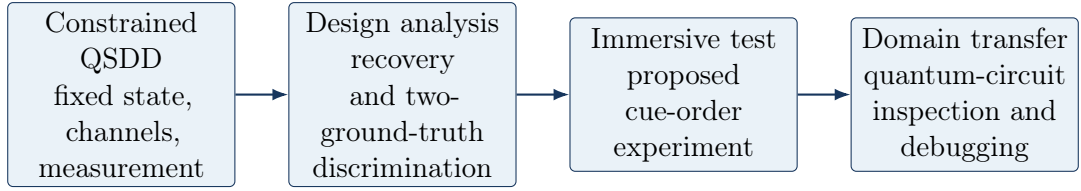

The paper makes five contributions:
\begin{enumerate}[leftmargin=1.6em]
  \item It separates physical measurement, quantum-like cognitive modeling, and interactional observation.
  \item It defines QSDD using consistent channel and instrument notation and derives a fixed-axis, five-parameter binary model in which the signs of both cue rotations are observable.
  \item It reports reproducible recovery and discrimination simulations under both QSDD and classical ground truths.
  \item It maps immersive telemetry to model evidence and specifies falsifiable hypotheses, classical baselines, and a counterbalanced VR protocol.
  \item It identifies a concrete transfer test in human--quantum computer interaction: representation-order effects during quantum-circuit inspection and debugging.
\end{enumerate}

\section{Observation and Collapse: An Epistemic Boundary}

\subsection{Physical measurement is not a settled theory of consciousness}

In textbook quantum mechanics, a state may be represented by a vector or density operator, unitary evolution preserves superposition, and measurement supplies probabilities for outcomes. Interpretations disagree about what the formalism says exists and whether an apparent collapse is physical, epistemic, relational, emergent through decoherence, or unnecessary. Historical proposals connecting mind and measurement make consciousness-centered collapse a legitimate topic in the philosophy of physics, but not an established empirical premise for cognitive science \citep{vonneumann1955foundations,wigner1961remarks,schlosshauer2005decoherence}.

Nothing in the framework below depends on a conscious observer physically selecting one branch of a microscopic wave function. The connection is structural rather than ontological: a context-sensitive state is transformed by an intervention, and a registered outcome changes the state relevant to subsequent predictions.

\subsection{Three claims that must not be conflated}

\begin{table}[htbp]
\centering
\caption{Three distinct claims hidden by casual uses of ``observation'' and ``collapse.'' QSDD adopts only the second and third.}
\label{tab:claims}
\begin{tabularx}{\linewidth}{>{\bfseries}p{0.17\linewidth} Y >{\raggedright\arraybackslash}p{0.20\linewidth}}
\toprule
Claim & Meaning & Status in this paper \\
\midrule
Physical & Conscious awareness is necessary to collapse a physical quantum state. & Neither assumed nor tested. \\
Cognitive-formal & A quantum probability space may represent context-sensitive uncertainty and sequential judgment. & Adopted as a candidate model. \\
Interactional & Looking, moving, querying, and manipulating can change the informational and cognitive context of a spatial decision. & Adopted and experimentally testable. \\
\bottomrule
\end{tabularx}
\end{table}

This separation is more than rhetorical caution. If a quantum-like model succeeds, it supports a probabilistic structure for the task; it does not establish neural quantum coherence. If it fails, that failure bears on the model and task, not on quantum mechanics. The proposal is therefore compatible with multiple physical interpretations and with ordinary neurocomputational implementation.

\subsection{Collapse as an operational term}

We use \emph{decision resolution} for the event at which a task records one alternative: crossing into a corridor, selecting a waypoint, grasping an object, or confirming a route. ``Collapse'' may be used as shorthand for the corresponding conditional state update, but it carries no claim of instantaneous, irreversible, or conscious physical collapse. Human commitment can be graded, delayed, revised, or reopened. A suitable formalism must therefore represent both sharp and unsharp measurements.

\section{Background}

\subsection{Spatial decision making is embodied and context dependent}

Spatial navigation integrates egocentric and allocentric information, landmarks, remembered routes, goals, and self-motion cues. These sources need not be simultaneously available or mutually consistent. Immersive experiments show that more than one spatial reference frame can support working memory after self-movement, with gaze providing a behavioral signature of which representation is recruited \citep{draschkow2022frames}. Multisensory input can also change navigation strategies and performance rather than simply add redundant evidence \citep{ruotolo2023multisensory}. Such findings motivate a state representation in which cue sequence and viewpoint may change the operative decision basis.

VR provides both ecological richness and experimental control \citep{huang2023leveraging}, but it is not transparent. Locomotion technique, field of view, display latency, embodiment, simulator familiarity, and missing vestibular or proprioceptive cues can alter behavior. Recent evidence that neural representations can differ between virtual and physical navigation reinforces the need to treat immersion as a modeled context rather than a neutral delivery channel \citep{yan2026divergent}.

\subsection{Immersive interaction produces process evidence}

Immersive analytics treats movement, spatial arrangement, manipulation, and viewpoint as components of sensemaking rather than interface overhead alone \citep{dwyer2018immersive,marriott2018immersive}. Work on embodied provenance shows that interaction histories can help reconstruct how an analysis unfolded \citep{zhang2023provenance}. Gaze and head motion can predict navigational difficulty or the need for aid \citep{alghofaili2019lost}, but any single channel is ambiguous: gaze may indicate interest, confusion, salience, or motor preparation. QSDD therefore uses multimodal sequences and task events, not gaze alone, to constrain a latent-state model.

QTerra applies this process orientation to quantum-circuit inspection by capturing gaze, movement, selection, parameter manipulation, cost-surface inspection, and revisitation \citep{basu2026qterra}. QSDD inherits the same methodological commitment: final accuracy is insufficient to distinguish how a person arrived at a decision.

\subsection{Quantum cognition as a probabilistic modeling program}

Classical probability places events in a common Boolean algebra and typically assumes a joint distribution for relevant variables. Quantum probability represents events with subspaces or positive operators on a Hilbert space. Incompatible questions need not share a single joint distribution, and sequential measurements may not commute. This structure has been used to model order effects, conjunction and disjunction errors, violations of the sure-thing principle, and dynamic judgment \citep{pothos2009quantum,busemeyer2011conjunction,pothos2013direction,wang2014order,yearsley2016tutorial}.

The strongest evidence for a quantum-like account is not that a post hoc interference term fits data. A flexible classical model can often do the same. Strong evidence requires an independently motivated state space, constrained operators, parameter recovery, novel prediction, and comparison with plausible classical models on held-out data \citep{pothos2022review}. The QQ equality is one example of a structural prediction derived from a quantum question-order model and tested across many surveys \citep{wang2014order}. QSDD adopts this predictive standard.

Context dependence must also be distinguished from contextuality in its stronger technical sense. A distribution can change across contexts because of direct influences without violating every noncontextual representation. Contextuality-by-Default provides one framework for separating those possibilities, and behavioral demonstrations require experimental systems designed for that analysis \citep{basieva2019contextuality}. The present paper predicts context-dependent and potentially noncommuting updates; it does not claim to have demonstrated contextuality. A dedicated cyclic design would be required for that stronger claim.

Quantum cognition also offers a computational path: cognitive models expressed in Hilbert space can be compiled into actual quantum circuits without implying that the biological brain uses qubits \citep{widdows2023circuits}. This makes it possible to move in stages from quantum-like behavioral modeling, to circuit representations of those models, and eventually to experiments on quantum simulators or hardware. The empirical HCI program and the computational quantum program can therefore converge without being conflated.

\subsection{Human--quantum computer interaction as a transfer domain}

Human--quantum computer interaction concerns the people, representations, tools, and practices through which quantum computation becomes usable \citep{ashktorab2019thinking}. Recent interface work has moved beyond agendas to concrete techniques for circuit composition, machine inspection, circuit comparison, optimization review, and result analysis \citep{kim2025toward}. Human-centered quantum software engineering similarly emphasizes practice, responsibility, and cooperative support \citep{bano2026human}. These developments motivate QTerra and the later transfer study in this paper, but they do not validate QSDD. The empirical connection must be earned by testing whether a model first constrained in navigation predicts behavior in quantum-circuit tasks.

\section{Quantum-Like Spatial Decision Dynamics}

\subsection{Decision state}

Let the task-relevant cognitive state at time $t$ be a density operator
\begin{equation}
  \rho_t \in \mathbb{C}^{d\times d}, \qquad
  \rho_t \succeq 0, \qquad \Tr(\rho_t)=1,
  \label{eq:density}
\end{equation}
on a finite-dimensional Hilbert space $\Hilb$. Density operators allow both a coherent pure state $\rho=\ket{\psi}\bra{\psi}$ and a mixed state representing population heterogeneity, unobserved context, or within-person uncertainty. The basis can encode route alternatives,
\begin{equation}
  \mathcal{B}_{R}=\{\ket{r_1},\ldots,\ket{r_K}\},
\end{equation}
or strategies such as landmark following, allocentric map use, exploration, and backtracking. These are modeling coordinates, not hypothesized neurons.

For a binary left-right choice, a minimal pure state is
\begin{equation}
  \ket{\psi_t}=\cos\!\left(\frac{\theta_t}{2}\right)\ket{L}
  +e^{i\phi_t}\sin\!\left(\frac{\theta_t}{2}\right)\ket{R}.
  \label{eq:qubit}
\end{equation}
The phase $\phi_t$ has no mystical interpretation; it parameterizes context-dependent interference that is invisible in a single route measurement but can affect predictions after a basis transformation.

\subsection{Perceptual and interactive contexts}

A spatial event $c_t$ can be represented as a completely positive trace-preserving map
\begin{equation}
  \rho_{t+1}^{-}=\Context_{c_t}(\rho_t)
  =\sum_j K_{j,c_t}\rho_t K_{j,c_t}^{\dagger},
  \qquad \sum_j K_{j,c_t}^{\dagger}K_{j,c_t}=I.
  \label{eq:contextmap}
\end{equation}
Examples include revealing a landmark, rotating the body, opening a map, hearing a verbal cue, manipulating an object, or receiving performance feedback. A unitary transformation is a special case. Non-unitary maps can represent noise, forgetting, distraction, or the loss of coherence between competing representations.

Context is coupled to action. The cue a user receives at $t+1$ may depend on the viewpoint or location chosen at $t$. QSDD therefore treats the observed sequence $c_{1:T}$ as an endogenous perception-action history when estimating causal effects. Randomization is needed to distinguish environmental manipulation from self-selection.

\subsection{Judgment and commitment as quantum instruments}

A generalized measurement has effects $E_y\succeq 0$ satisfying $\sum_y E_y=I$. The probability of outcome $y$ is
\begin{equation}
  \Prob(y_t=y\mid \rho_t,c_t)=\Tr(E_{y,c_t}\rho_t).
  \label{eq:born}
\end{equation}
To specify both outcome probability and post-outcome state, use a quantum instrument $\{\Instrument_y\}$:
\begin{equation}
  p_y=\Tr[\Instrument_y(\rho)], \qquad
  \rho'_{y}=\frac{\Instrument_y(\rho)}{p_y}.
  \label{eq:instrument}
\end{equation}
An explicit route choice can be modeled with projectors $\projector{r_i}=\ket{r_i}\bra{r_i}$. A brief glance, confidence probe, or incomplete turn can instead be represented with an unsharp positive-operator-valued measure (POVM). This matters because immersive behavior rarely supports a single instantaneous commitment point.

\begin{definition}[Operational commitment]
For a preregistered task, commitment occurs at the first event satisfying a behavioral criterion that makes an alternative externally registered, such as crossing a decision boundary for at least $\tau$ seconds without reversal. The criterion is defined independently of the fitted QSDD model.
\end{definition}

\subsection{Order effects and noncommutativity}

Let $\Context_L$ denote a landmark-cue channel and $\Context_M$ an allocentric map-cue channel. They are noncommuting as superoperators when
\begin{equation}
  \Context_M\!\circ\!\Context_L\neq \Context_L\!\circ\!\Context_M.
  \label{eq:commutator}
\end{equation}
For the same final route effect $E_d$, the two sequence probabilities are
\begin{align}
  p(d\mid L\!\rightarrow\!M)
  &=\Tr\!\left[E_d\,\Context_M\!\left(\Context_L(\rho_0)\right)\right],\\
  p(d\mid M\!\rightarrow\!L)
  &=\Tr\!\left[E_d\,\Context_L\!\left(\Context_M(\rho_0)\right)\right].
  \label{eq:order}
\end{align}
These expressions describe unobserved cue exposures. If $L$ or $M$ instead denotes a registered intermediate outcome, the corresponding quantum instrument from \cref{eq:instrument} must be used and the post-outcome state normalized. This distinction prevents a cue, a measurement, and an observed result from being represented by the same ambiguous symbol.

The empirical target is the order contrast
\begin{equation}
  \Delta_{LM}^{(d)}=p(d\mid L\!\rightarrow\!M)-p(d\mid M\!\rightarrow\!L).
  \label{eq:delta}
\end{equation}
An order effect alone is not uniquely quantum; memory, anchoring, primacy, recency, or asymmetric noise can generate one classically. The quantum-like claim earns support only if one constrained operator model predicts a family of order effects, transition patterns, and choice distributions better than those alternatives.

\begin{proposition}[Sequence sensitivity]
If $\Context_M\circ\Context_L=\Context_L\circ\Context_M$ and the same initial state and final effect are used, then $\Delta_{LM}^{(d)}=0$. A reliable order effect therefore rejects that commuting QSDD model, but it does not by itself reject classical models with history-dependent state transitions.
\end{proposition}

\subsection{Interference and the law of total probability}

Suppose $A$ denotes a spatial hypothesis, such as ``the landmark-defined route is reliable,'' and $D$ a route decision. Let $U$ denote a protocol in which the alternatives $A$ and $\neg A$ remain unresolved, and let $Q$ denote a matched protocol that queries or reveals the partition before the decision. Under the classical mixture and a cross-protocol stability assumption,
\begin{equation}
  p_U(D)=p_Q(D\mid A)p_Q(A)+p_Q(D\mid \neg A)p_Q(\neg A).
  \label{eq:ltp}
\end{equation}
The empirical interference residual is therefore
\begin{equation}
 I_D=p_U(D)-\left[p_Q(D\mid A)p_Q(A)+p_Q(D\mid \neg A)p_Q(\neg A)\right].
 \label{eq:interference-residual}
\end{equation}
In a two-path quantum-like construction this residual can be parameterized as
\begin{align}
 I_D=&\;2\sqrt{p_Q(D\mid A)p_Q(A)
 p_Q(D\mid \neg A)p_Q(\neg A)}\cos\varphi.
 \label{eq:interference}
\end{align}
The equality in \cref{eq:ltp} is not licensed if querying $A$ changes the meaning, sampling process, or population defining $D$. The protocols must therefore be matched, the stability assumption stated, and probe reactivity estimated. A nonzero $I_D$ is a descriptive interference-like quantity, not uniquely quantum evidence. Support requires that a phase estimated in training environments predict $I_D$ in held-out environments better than classical context and memory models.

\subsection{Progressive commitment and reopening}

The phenomenology of a route decision is often gradual. Let $p_i(t)=\Tr(\projector{r_i}\rho_t)$ and define decision-basis uncertainty as
\begin{equation}
  H_R(t)=-\sum_i p_i(t)\log p_i(t).
  \label{eq:entropy}
\end{equation}
$H_R$ is Shannon entropy over route outcomes in a specified basis. It is distinct from the basis-independent von Neumann entropy $S(\rho)=-\Tr(\rho\log\rho)$. A decisive cue can reduce $H_R$ even while $S(\rho)$ is unchanged under unitary evolution. This distinction prevents the loose claim that any reduction in behavioral uncertainty is physical decoherence.

New evidence can reopen a resolved or nearly resolved decision:
\begin{equation}
  \rho_{A}\xrightarrow{\Context_{\mathrm{conflict}}}
  \rho_{A/B}, \qquad H_R(\rho_{A/B})>H_R(\rho_A).
  \label{eq:reopen}
\end{equation}
Observable signatures may include head reorientation, return to a prior cue, path reversal, renewed map inspection, or increased response latency. These signatures constrain, but do not directly equal, latent uncertainty.

\subsection{Embodiment changes the measurement context}

In desktop tasks, context is often defined by stimulus content and question order. In immersion, context also includes pose, reachable action, scale, optic flow, proprioception, and the spatial cost of seeking information. We represent the effective context as
\begin{equation}
  c_t=(s_t,v_t,a_{t-1},m_t,g_t),
\end{equation}
where $s_t$ is scene state, $v_t$ viewpoint/body state, $a_{t-1}$ prior action, $m_t$ memory/task history, and $g_t$ interface or guidance state. ``The same cue'' viewed from a different pose need not instantiate the same operator.

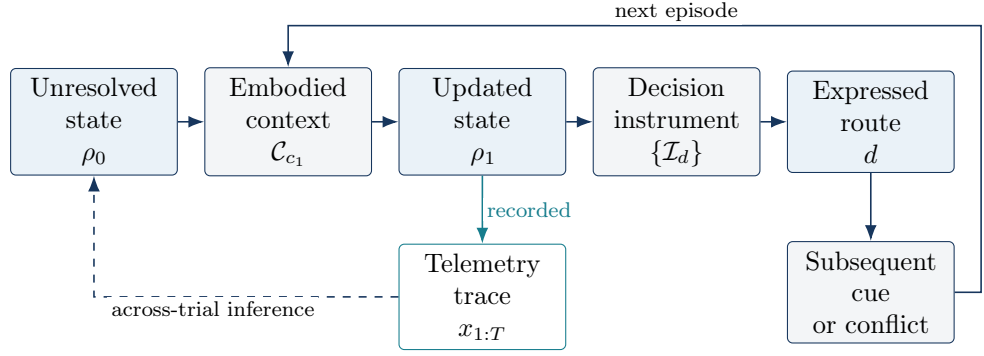
\begin{figure}[htbp]
\centering
\begin{tikzpicture}[
  node distance=0.9cm and 0.35cm,
  state/.style={draw=deepblue,rounded corners=2pt,fill=softblue,
    align=center,minimum height=1.05cm,text width=1.95cm,font=\small},
  operation/.style={draw=deepblue,rounded corners=2pt,fill=softgray,
    align=center,minimum height=1.05cm,text width=1.95cm,font=\small},
  outcome/.style={draw=deepblue,rounded corners=2pt,fill=softblue,
    align=center,minimum height=1.05cm,text width=1.95cm,font=\small},
  observed/.style={draw=teal,rounded corners=2pt,fill=white,
    align=center,minimum height=0.9cm,text width=1.95cm,font=\small},
  flow/.style={-{Latex[length=2.1mm]},semithick,draw=deepblue},
  emission/.style={-{Latex[length=2.1mm]},semithick,draw=teal},
  inference/.style={-{Latex[length=2.1mm]},semithick,dashed,draw=deepblue},
  edge label/.style={font=\scriptsize,fill=white,inner sep=1.5pt}
]
\node[state] (state0) {Unresolved\\state\\$\rho_0$};
\node[operation,right=of state0] (context1) {Embodied\\context\\$\Context_{c_1}$};
\node[state,right=of context1] (state1) {Updated\\state\\$\rho_1$};
\node[operation,right=of state1] (measure) {Decision\\instrument\\$\{\Instrument_d\}$};
\node[outcome,right=of measure] (action) {Expressed\\route\\$d$};
\node[observed,below=of state1] (trace) {Telemetry trace\\$x_{1:T}$};
\node[operation,below=of action] (next) {Subsequent cue\\or conflict};

\draw[flow] (state0)--(context1);
\draw[flow] (context1)--(state1);
\draw[flow] (state1)--(measure);
\draw[flow] (measure)--(action);
\draw[emission] (state1)--node[edge label,right,text=teal]{recorded}(trace);
\draw[flow] (action)--(next);
\draw[flow] (next.east)--++(0.35,0)
  |- node[edge label,pos=0.72,above]{next episode}
  ($(context1.north)+(0,0.55)$)--(context1.north);
\draw[inference] (trace.west)--++(-0.45,0)
  -| node[edge label,pos=0.28,below]{across-trial inference}(state0.south);
\end{tikzpicture}
\caption{QSDD perception--action and inference loop. Solid navy arrows show the behavioral sequence within and between episodes; the teal branch marks recorded telemetry. Telemetry is evidence about state evolution, not a direct reading of the cognitive state. The dashed arrow denotes model inference across repeated trials rather than literal feedback from a data log to the participant.}
\label{fig:framework}
\end{figure}

\section{A Constrained Binary Model}

The general framework is deliberately expressive; a confirmatory test requires a much smaller model. For a binary route choice, represent the state by a Bloch vector and fix the decision basis before observing confirmatory data. The initial state is
\begin{equation}
 \begin{aligned}
 \bm r_0
 &=\left(\sin\theta_0\cos\phi_{\!*},
          \sin\theta_0\sin\phi_{\!*},
          \cos\theta_0\right)^{\mathsf T},
 &\phi_{\!*}&=\frac{\pi}{4},\\
 \rho_0&=\tfrac12\left(I+\bm r_0\cdot\bm\sigma\right).&&
 \end{aligned}
 \label{eq:bloch-initial}
\end{equation}
where $\ket{L}$ and $\ket{R}$ are the north and south poles, respectively. The azimuth $\phi_{\!*}$ is a fixed geometric constant, not an estimated parameter. Choosing both $\sin\phi_{\!*}\neq0$ and $\cos\phi_{\!*}\neq0$ prevents either cue sign from disappearing under route-basis measurement. Environment $e$ supplies preregistered, signed cue strengths $\ell_e,m_e\in[-1,1]$. The landmark cue rotates about the $y$ axis and the map cue about the $x$ axis:
\begin{equation}
 R_L(\ell_e)=R_y(\omega_L\ell_e),\qquad
 R_M(m_e)=R_x(\omega_Mm_e).
 \label{eq:fixed-rotations}
\end{equation}
Between cue operations, dephasing is represented in Bloch coordinates by $D_\kappa=\operatorname{diag}(\kappa,\kappa,1)$. The two predicted terminal states are
\begin{align}
 \bm r_{LM,e}&=D_\kappa R_M(m_e)D_\kappa R_L(\ell_e)\bm r_0,\\
 \bm r_{ML,e}&=D_\kappa R_L(\ell_e)D_\kappa R_M(m_e)\bm r_0.
 \label{eq:sequence-bloch}
\end{align}
The route probability, including a symmetric lapse rate $\epsilon$, is
\begin{equation}
 p(R\mid \bm r)=\frac{\epsilon}{2}+(1-\epsilon)\frac{1-r_z}{2}.
 \label{eq:binary-probability}
\end{equation}
To verify signed-cue observability, write $a_e=\omega_L\ell_e$, $b_e=\omega_Mm_e$, $s_0=\sin\theta_0$, and $c_0=\cos\theta_0$. The two route-basis components are
\begin{equation}
\begin{aligned}
 r_{z,LM,e}
 &=\left(c_0\cos a_e-s_0\cos\phi_{\!*}\sin a_e\right)\cos b_e
   +\kappa s_0\sin\phi_{\!*}\sin b_e,\\
 r_{z,ML,e}
 &=-\kappa s_0\cos\phi_{\!*}\sin a_e
   +\left(s_0\sin\phi_{\!*}\sin b_e+c_0\cos b_e\right)\cos a_e.
\end{aligned}
\label{eq:signed-observability}
\end{equation}
Each expression contains a term odd in $b_e$, and each sequence retains a term odd in $a_e$ under the stated parameter bounds. Thus reversing either signed cue generically changes the predicted route probability. The equality can persist at isolated parameter or cue values, but it is no longer a structural $m_e\leftrightarrow-m_e$ symmetry.

The complete population-level parameter vector is therefore
\begin{equation}
 \Theta=(\theta_0,\omega_L,\omega_M,\kappa,\epsilon).
 \label{eq:parameter-vector}
\end{equation}
One $\Theta$ must predict both cue orders and every environment. The model is not permitted an environment-specific phase, azimuth, or order coefficient. Its empirical content comes from that shared nonlinear surface: after $\Theta$ is estimated on training environments, the sign and magnitude of the order contrast in a new $(\ell_e,m_e)$ pair are fixed.

This parameterization also fixes otherwise consequential gauge freedoms. The route basis, rotation axes, initial azimuth, cue-strength scale, and parameter bounds are declared in advance. Sign conventions are fixed by practice-trial cue validity. Hierarchical extensions may allow participant variation around $\Theta$, but should preserve the axes and fixed azimuth and use shrinkage rather than cell-specific operators. Catch trials are needed to identify $\epsilon$ separately from weak cue response and dephasing.

\subsection{A prespecified classical logistic comparator}

The minimal classical comparator is not a cue-independent mixture. It is a logistic model with main effects, cue interaction, order, and order-by-cue interactions:
\begin{equation}
 \operatorname{logit}p(R)=\beta_0+\beta_L\ell+\beta_Mm+\beta_{LM}\ell m
 +\beta_Oo+\beta_{OL}o\ell+\beta_{OM}om,
 \label{eq:classical-logistic}
\end{equation}
where $o\in\{-1,+1\}$ encodes sequence. This seven-parameter model can express primacy, recency, and cue-dependent order effects. Later confirmatory analyses should also include classical latent-state and sequential-sampling models; \cref{eq:classical-logistic} is used below because it is more flexible in parameter count than the five-parameter QSDD model while remaining estimable in a compact design analysis.

\section{Simulation-Based Recovery and Model Discrimination}
\label{sec:simulation}

\subsection{Design and estimands}

We conducted a reproducible simulation to answer two pre-empirical questions: whether the constrained QSDD parameters can be recovered, and whether held-out prediction selects the correct model family when either QSDD or the classical model generates the data. This is a computational stress test of the proposed analysis, not evidence about human cognition.

Each replicate used 12 training environments and six held-out environments. Cue-strength pairs were continuously sampled with stratification over $[-1,1]^2$, and both cue orders were observed in every environment. We evaluated $N\in\{40,80,120,240,480\}$ Bernoulli observations per environment-order cell and generated 40 replicates for each combination of $N$ and ground truth. The fixed initial azimuth was $\phi_{\!*}=\pi/4$, and the QSDD generating parameters were
\begin{equation}
 \Theta^*=(1.35,1.10,0.95,0.86,0.035),
\end{equation}
and the classical generating vector for \cref{eq:classical-logistic} was
\begin{equation}
 \bm\beta^*=(-0.15,1.10,-0.95,0.65,0.45,-0.55,0.40).
\end{equation}
Both models were estimated by bounded maximum likelihood with multiple starting points. Performance was evaluated by held-out log predictive density per Bernoulli observation,
\begin{equation}
 \Delta\mathrm{ELPD}/n_{\mathrm{test}}
 =\frac{\log p_{\mathrm{QSDD}}(y_{\mathrm{test}})-
 \log p_{\mathrm{classical}}(y_{\mathrm{test}})}{n_{\mathrm{test}}},
 \label{eq:simulation-elpd}
\end{equation}
and by normalized QSDD parameter RMSE. The random seed was 20261002. The script, replicate-level output, summary table, and metadata are included with the source package.

This experiment examines one prespecified generating vector for each model family rather than a parameter-wide robustness sweep. Its conclusions are therefore local to the stated design settings; broader claims require sensitivity analyses over parameter regimes, participant heterogeneity, and trial dependence.

\begin{figure}[htbp]
\centering
\includegraphics[width=0.96\linewidth]{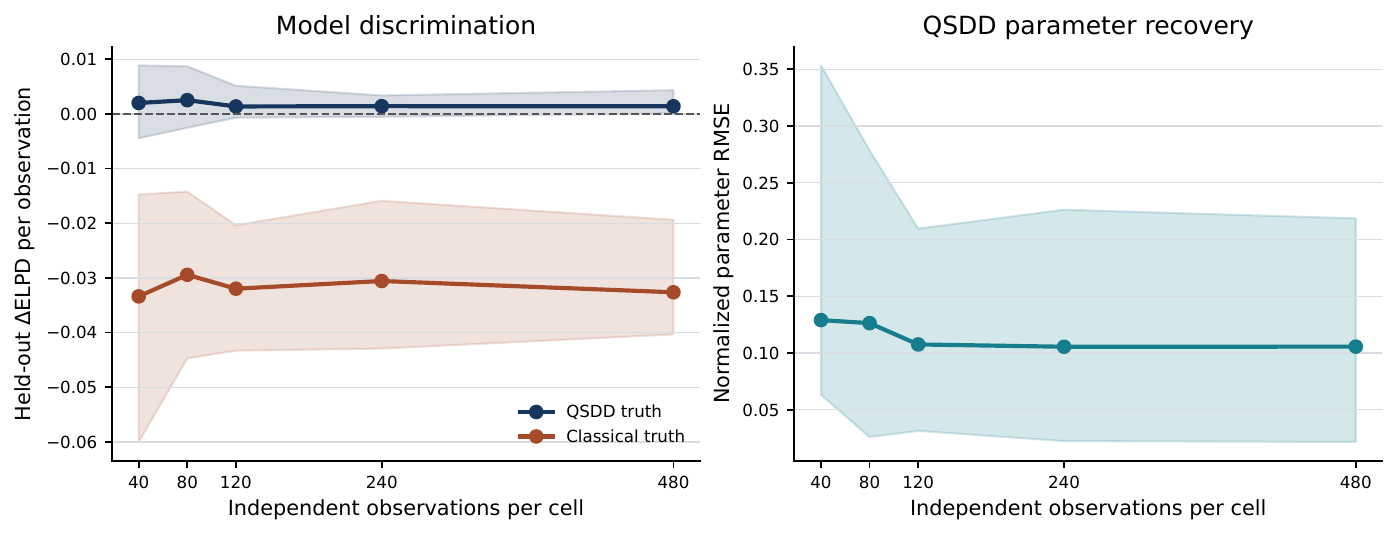}
\caption{Synthetic recovery and discrimination results. Lines show medians across 40 replicates; bands show the 5th--95th percentile interval. Positive $\Delta$ELPD favors QSDD. These curves characterize the simulated design and effect sizes, not expected human-study outcomes.}
\label{fig:simulation}
\end{figure}

\begin{table}[htbp]
\centering
\caption{Simulation summary. Win rate is the proportion of replicates with positive held-out $\Delta$ELPD.}
\label{tab:simulation}
\begin{tabular}{llrrrr}
\toprule
Truth & $N$ & Median $\Delta$ELPD/obs. & 5th pct. & 95th pct. & QSDD win rate \\
\midrule
QSDD & 40  & 0.0020 & -0.0045 & 0.0089 & 0.75 \\
QSDD & 80  & 0.0025 & -0.0025 & 0.0087 & 0.83 \\
QSDD & 120 & 0.0014 & -0.0007 & 0.0051 & 0.83 \\
QSDD & 240 & 0.0014 & -0.0005 & 0.0034 & 0.85 \\
QSDD & 480 & 0.0014 &  0.0001 & 0.0043 & 0.95 \\
\addlinespace
Classical & 40  & -0.0334 & -0.0600 & -0.0147 & 0.00 \\
Classical & 80  & -0.0294 & -0.0447 & -0.0142 & 0.00 \\
Classical & 120 & -0.0320 & -0.0433 & -0.0204 & 0.00 \\
Classical & 240 & -0.0306 & -0.0429 & -0.0159 & 0.00 \\
Classical & 480 & -0.0326 & -0.0403 & -0.0194 & 0.00 \\
\bottomrule
\end{tabular}
\end{table}

\subsection{What the simulation establishes---and what it does not}

Under the selected QSDD ground truth, the win rate was 0.75 at $N=40$, 0.85 at $N=240$, and 0.95 at $N=480$, where the 5th percentile of $\Delta$ELPD was positive. Under the classical ground truth, QSDD was never preferred in the 200 simulated datasets spanning all five sample sizes. The comparison procedure therefore did not manufacture a QSDD advantage merely because the quantum-like model was included. The corrected signed-cue geometry also shows that $N=240$ is less decisive than the earlier symmetry-restricted parameterization suggested.

Median normalized parameter RMSE under QSDD ground truth was 0.129 at $N=40$ and 0.106 at $N=480$. At $N=480$, the medians for $(\theta_0,\omega_L,\omega_M,\kappa)$ were $(1.351,1.119,0.960,0.855)$, close to their generating values. The lapse median was 0.035, equal to the generating value to the reported precision, but its 5th--95th percentile interval remained wide at $[0,0.106]$. This identifies a design weakness: confirmatory estimation should either constrain lapse using independent catch trials or omit it when such trials are unavailable.

The result does not show that people follow QSDD, and $N=480$ is not a universal sample-size prescription. The simulated effect sizes, lack of participant heterogeneity, balanced cells, fixed initial azimuth, and known cue-strength scale are optimistic relative to a real VR study. Pilot-based design simulation must therefore replace these illustrative values before recruitment. The value of the present result is narrower but important: the repaired model is executable, both cue signs affect prediction, its central parameters are recoverable under stated conditions, and the same analysis can reject it when the ground truth is classical.

\section{From Immersive Telemetry to Model Evidence}

Telemetry provides observations $x_t$ generated conditionally by a latent decision state and the task context:
\begin{equation}
  x_t \sim p(x_t\mid \rho_t,c_t,\eta),
  \label{eq:emission}
\end{equation}
where $\eta$ contains participant- and device-level parameters. The framework does not label a gaze fixation as a measurement merely because it was recorded. An event is assigned theoretical status by experimental design and model specification.

\begin{table}[htbp]
\centering
\caption{Operational mapping from immersive observations to theoretical roles. Each mapping is a hypothesis subject to validation.}
\label{tab:mapping}
\begin{tabularx}{\linewidth}{>{\bfseries}p{0.19\linewidth} Y Y}
\toprule
Observed event & Candidate theoretical role & Principal ambiguity or control \\
\midrule
Landmark fixation & Exposure to a landmark context & Peripheral perception; salience without use \\
Map opening & Self-initiated information measurement & Confounded with prior uncertainty \\
Head/body rotation & Basis or viewpoint transformation & Motor convenience and tracking noise \\
Pause or hesitation & Sustained unresolved competition & Reading time, fatigue, scene complexity \\
Boundary crossing & Operational route commitment & Accidental crossing; enforce dwell criterion \\
Backtracking & Reopening after conflicting evidence & Motor error or exploration policy \\
Repeated inspection & Evidence re-sampling or low confidence & Interface discoverability and novelty \\
Confidence report & Explicit, potentially state-changing probe & Probe reactivity; counterbalance or omit \\
\bottomrule
\end{tabularx}
\end{table}

A practical telemetry schema should synchronize: trial and condition identifiers; cue onset and offset; head, hand, and body poses; gaze ray and validity; scene-object intersections; map or panel events; movement speed and angular velocity; path coordinates; decision-boundary crossings; reversals; response time; confidence; and task outcome. Sampling rates should be recorded rather than assumed, and derived features should preserve a link to raw timestamps.

\section{A Preregisterable Immersive Experiment}

\subsection{Research question}

\textbf{RQ:} When people integrate landmark and allocentric map information during immersive navigation, does the sequence and mode of observation change the decision state in ways better predicted by a constrained quantum-like model than by classical alternatives?

\subsection{Task and apparatus}

Participants navigate a family of matched virtual environments containing repeated decision nodes. At each critical node, two cue families provide route-relevant evidence:
\begin{itemize}
  \item \textbf{Landmark cue ($L$):} an egocentric, scene-embedded object or sign associated with a route.
  \item \textbf{Map cue ($M$):} an allocentric overhead representation indicating layout or goal relation.
\end{itemize}
The environment is designed so that cue reliability is learned in practice trials and manipulated in the experiment. Critical trials include congruent cues, conflicting cues, single-cue trials, and unresolved/no-query trials. The same geometry is mirrored and texture-varied to reduce memorization.

A head-mounted display with eye tracking records gaze, head pose, hand pose, and locomotion. A matched desktop condition may be included as a secondary factor, but the core quantum-like test is within immersive VR. Physical walking is preferred where space permits; if controller locomotion is used, it is held constant and modeled.

\subsection{Experimental factors}

The minimal within-participant design crosses:
\begin{enumerate}
  \item \textbf{Cue order:} $L\rightarrow M$ versus $M\rightarrow L$.
  \item \textbf{Cue relation:} congruent versus conflicting.
  \item \textbf{Observation mode:} forced sequential exposure versus participant-controlled query.
  \item \textbf{Decision probe:} no intermediate probe versus a preregistered unsharp confidence probe.
\end{enumerate}
Single-cue and unresolved trials estimate terms needed for the total-probability test. Trial order, route side, environment identity, cue reliability, and goal position are counterbalanced. Forced-order trials identify causal sequence effects; self-directed trials test ecological generalization without being used as the primary causal estimate.

\Needspace{0.34\textheight}
\subsection{Trial sequence}

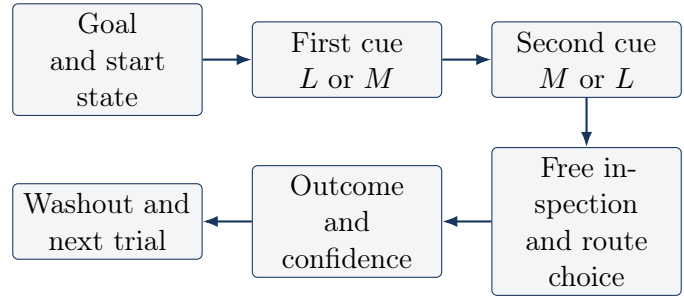
\begin{figure}[H]
\centering
\begin{tikzpicture}[
  node distance=0.65cm,
  trialstep/.style={draw=deepblue,rounded corners=2pt,fill=softgray,align=center,minimum height=0.75cm,text width=2.25cm},
  arr/.style={-{Latex[length=2mm]},thick,draw=deepblue}
]
\node[trialstep] (start) {Goal and start\\state};
\node[trialstep,right=of start] (cue1) {First cue\\$L$ or $M$};
\node[trialstep,right=of cue1] (cue2) {Second cue\\$M$ or $L$};
\node[trialstep,below=of cue2] (choice) {Free inspection\\and route choice};
\node[trialstep,left=of choice] (outcome) {Outcome and\\confidence};
\node[trialstep,left=of outcome] (reset) {Washout and\\next trial};
\draw[arr] (start)--(cue1);
\draw[arr] (cue1)--(cue2);
\draw[arr] (cue2)--(choice);
\draw[arr] (choice)--(outcome);
\draw[arr] (outcome)--(reset);
\end{tikzpicture}
\caption{Core forced-order trial. Exposure duration and locomotor distance between cues are matched across orders. Intermediate probes appear only in designated blocks because asking a question may itself change the state.}
\label{fig:trial}
\end{figure}

\subsection{Confirmatory hypotheses to preregister}

\textbf{H1: Frozen-operator order prediction.} On matched conflict trials,
\begin{equation}
  \Prob(R\mid L\!\rightarrow\!M)\neq \Prob(R\mid M\!\rightarrow\!L),
\end{equation}
with both direction and magnitude predicted from pilot-estimated $\Theta$ and frozen before confirmatory data collection. A significant omnibus order coefficient without successful cross-environment prediction does not support QSDD.

\textbf{H2: Shared nonlinear surface.} A single $\Theta$ will predict the order-contrast surface across preregistered $(\ell_e,m_e)$ pairs. Environment-specific phases, bases, or order parameters are prohibited in the confirmatory model.

\textbf{H3: Protocol-qualified interference.} After checking the cross-protocol stability assumptions surrounding \cref{eq:ltp}, the unresolved condition will exhibit a nonzero $I_D$ from \cref{eq:interference-residual}; a phase estimated on training environments must predict its sign and magnitude in held-out environments.

\textbf{H4: Progressive commitment.} On preregistered diagnostic-cue trials, the within-trial change $\Delta H_R=H_R(t_{\mathrm{post}})-H_R(t_{\mathrm{pre}})$ is predicted to be negative on average, and lower post-cue $H_R$ is predicted to precede shorter time to operational commitment. The primary estimand is the participant-level mean $\Delta H_R$; inference will use a hierarchical model controlling for path distance, cue visibility, and motor speed. The confirmatory sign, time window, and smallest effect of interest will be frozen after the pilot and before confirmatory collection.

\textbf{H5: Reopening.} A late conflicting cue will increase route-basis uncertainty and the probability of a preregistered behavioral reversal signature relative to a matched confirming cue.

\textbf{H6: Cross-medium transportability.} If a desktop comparison is included, frozen desktop operators will be transferred to VR and compared with a medium-specific hierarchical model. Transport failure will be declared only if the medium-specific model improves held-out ELPD beyond a preregistered threshold and at least one operator shift falls outside its equivalence region. The threshold and equivalence regions will be frozen after the pilot. This tests parameter transport, not whether VR universally improves accuracy.

\subsection{Outcome measures}

Primary outcomes are route choice and sequence-conditioned choice probability. Secondary outcomes are decision time, path efficiency, confidence, cue-query order, gaze transitions among predefined regions of interest, pause duration, angular reorientation, boundary reversals, backtracking distance, and number of cue revisits. Simulator discomfort, VR familiarity, gaming experience, spatial ability, and learned cue reliability are recorded as covariates or stratification variables defined in advance.

\subsection{Model set and comparison}

The confirmatory analysis should compare models that receive the same exogenous inputs:
\begin{table}[htbp]
\centering
\caption{Required model comparisons. The QSDD model is not privileged by the analysis plan.}
\label{tab:models}
\begin{tabularx}{\linewidth}{>{\bfseries}p{0.22\linewidth} Y Y}
\toprule
Model family & What it captures & Key comparison \\
\midrule
Hierarchical logistic/Bayesian & Cue weights, interactions, primacy/recency, participant variation & Strong descriptive baseline for final choice \\
Hidden Markov or POMDP & Latent strategies, uncertainty, action-dependent observation & Classical sequential state alternative \\
Noncommutative classical transition & Order-sensitive matrix or nonlinear state update without Hilbert-space constraints & Tests whether sequence sensitivity alone explains the result \\
Drift-diffusion / race & Time-varying evidence and commitment thresholds & Joint choice and response-time alternative \\
Reinforcement learning & Learned cue values and exploration policy & Practice and reliability learning \\
QSDD & Noncommuting contexts, generalized measurement, interference, state update & Constrained prediction across order and unresolved conditions \\
\bottomrule
\end{tabularx}
\end{table}

These alternatives are grounded in established classical accounts of hidden-state inference and partially observable control \citep{rabiner1989tutorial,kaelbling1998pomdp}, sequential sampling \citep{ratcliff2016diffusion}, and learned value updating \citep{sutton2018reinforcement}. Their confirmatory implementations must be specified with the same inputs, train/test partitions, and prior-sensitivity checks used for QSDD.

Evaluation uses participant-stratified held-out log predictive density, calibration, posterior predictive checks, and information criteria or marginal likelihood where assumptions permit. Parameter counts alone are insufficient; effective complexity, hierarchical shrinkage, and prior sensitivity are reported. A model must recover known parameters in simulation before its fitted latent states are interpreted. Ablations remove phase, noncommutativity, or telemetry emissions to identify which component contributes predictive value.

The decisive comparison is predictive, not terminological:
\begin{equation}
  \Delta_{\mathrm{ELPD}}
  =\mathrm{ELPD}_{\mathrm{QSDD}}-\max_m \mathrm{ELPD}_{m,\mathrm{classical}}.
  \label{eq:elpd}
\end{equation}
Support requires a positive, uncertainty-qualified $\Delta_{\mathrm{ELPD}}$ on held-out participants and environments, together with successful parameter recovery and preregistered qualitative predictions.

\subsection{Sample size and uncertainty}

The final sample size should be determined by simulation-based design analysis, not a generic rule. The illustrative simulation in \cref{sec:simulation} shows why: after repairing signed-cue observability, QSDD was selected in 85\% of its own generating datasets at $N=240$ and reached 95\% only at $N=480$ under the stated assumptions. Pilot data must supply plausible participant heterogeneity, order contrast, lapse, repeated-measure dependence, and missing-gaze patterns. The selected participant count and trials per participant should achieve a preregistered probability of recovering the primary order surface and discriminating the candidate models under both QSDD and classical ground truths. Exclusions, stopping rules, missing-gaze handling, and the confirmatory/exploratory boundary are specified before collection.

\subsection{Threats to inference}

Several apparent quantum-like signatures have mundane explanations. Unequal cue duration can mimic order effects; a confidence question can create the effect it measures; gaze-contingent rendering can couple measurement and stimulus; scene salience can masquerade as preference; repeated trials can induce reinforcement learning; and simulator discomfort can alter search behavior. The study should therefore use matched exposure, manipulation checks, neutral probe wording, calibration trials, counterbalancing, and hierarchical participant effects. A no-probe group is essential for estimating probe reactivity.

\section{Transfer to Human--Quantum Computer Interaction}

The navigation experiment isolates the cognitive model from quantum-domain knowledge and interface novelty. A separate transfer test can then ask whether the constrained process model generalizes to a genuinely quantum-computing task. QTerra provides one such testbed by treating a variational circuit as an immersive work object \citep{basu2026qterra}. A user may inspect circuit structure, approach a gate, manipulate a parameter, view a local cost-landscape slice, revisit an ansatz layer, and then decide where an error lies. The alternatives are no longer literal left and right turns but competing debugging hypotheses or next actions.

For example, define $C$ as circuit-structure inspection, $S$ as cost-surface inspection, and $D$ as an error-localization decision. QSDD predicts that
\begin{equation}
  \Prob(D\mid C\!\rightarrow\!S)
  \quad\text{may differ from}\quad
  \Prob(D\mid S\!\rightarrow\!C)
\end{equation}
when the two representations induce incompatible judgment contexts. QTerra's telemetry can test whether the order affects not only the final answer but also subsequent revisits, parameter edits, and confidence. This extension closes a distinctive loop: humans reason about quantum circuits, while quantum-probability models are evaluated as accounts of how that reasoning unfolds.

The transfer test should preserve the inferential discipline of the navigation study. Circuit and surface views must be matched for exposure and information; representation order must be randomized; quantum expertise must be measured; and one operator family must predict held-out circuits. The primary outcome is not whether immersion is engaging, but whether representation order changes error localization, subsequent revisits, parameter edits, and confidence in the constrained pattern predicted by QSDD.

The transfer has three analytically separate layers:
\begin{enumerate}[leftmargin=1.6em]
  \item \textbf{Object layer:} quantum circuits, parameters, measurements, and optimization landscapes become objects of human inspection and action.
  \item \textbf{Cognitive layer:} QSDD models how representation order, observation, and interaction may transform the user's decision state.
  \item \textbf{Infrastructure layer:} telemetry, replay, provenance, and privacy-preserving computation make the reasoning process measurable and governable.
\end{enumerate}
Together, these layers define a testable contribution to human-centered quantum computing. They also identify where collaboration is required: quantum specialists validate task semantics, mathematical psychologists audit identifiability, and HCI researchers evaluate the representation and workflow.

\section{Scope and Collaboration Boundaries}

The phrase \enquote{quantum research} covers distinct communities. The primary locus of QSDD is the intersection of HCI, mathematical psychology, and human-centered quantum software. Quantum foundations motivate conceptual distinctions, but this paper does not propose an interpretation of quantum mechanics. Quantum hardware may eventually execute selected cognitive-model circuits, but no hardware advantage is claimed here.

\begin{table}[H]
\centering
\caption{Claim boundaries and collaboration requirements.}
\label{tab:positioning}
\begin{tabularx}{\linewidth}{>{\bfseries}p{0.19\linewidth} Y Y}
\toprule
Research layer & Contribution tested here or in the proposed program & Boundary or collaboration need \\
\midrule
Quantum foundations & Philosophical precision about observer, measurement, context, and interpretation & No claim to resolve the measurement problem \\
Quantum cognition & QSDD formalism, immersive order-effect experiments, telemetry-informed state models & Mathematical psychology and model-identifiability collaboration \\
Human-quantum interaction & Immersive representations, debugging workflows, user studies, gaze and interaction analysis & Quantum-domain task validation and expert participants \\
Quantum software and education & Tools for comprehension, circuit inspection, learning, and reflective replay & Integration with established quantum SDKs and curricula \\
Secure telemetry systems & Privacy, provenance, and secure processing of sensitive interaction traces & Cryptographic validation and realistic performance analysis \\
Human--AI--quantum systems & Adaptive decision support joining human judgment, AI mediation, and quantum workflows & Longer-term systems integration and governance \\
\bottomrule
\end{tabularx}
\end{table}

The boundary is deliberate. Evidence that QSDD predicts human choices would support a quantum-probability model of a task, not a quantum brain. Evidence that an immersive quantum interface improves debugging would support an HCI contribution, not a quantum algorithm. A later circuit implementation of a supported cognitive model would establish computational correspondence, not biological implementation. Keeping these claims separate is necessary for cumulative evidence.

\section{Privacy, Ethics, and Reproducibility}

Immersive telemetry can be identifying. Head and hand trajectories, gaze patterns, reaction times, and navigation strategies may function as behavioral signatures. Data minimization should begin at study design: collect only channels required by a hypothesis; separate identity from trial data; document sampling rates and transformations; restrict raw replay access; and state retention and sharing policies during consent.

Where cross-institutional or cloud analysis is necessary, aggregate release, differential privacy, secure enclaves, or encrypted computation may reduce exposure. A quantum-resilient homomorphic architecture is a long-term systems direction for analytics over sensitive motion and gaze data, but its cost, leakage model, and utility must be evaluated rather than assumed \citep{basu2026he}. Privacy technology does not replace informed consent or governance.

Reproducibility requires more than releasing a Unity scene. A complete research object should include environment versions, cue randomization seeds, coordinate conventions, event schema, gaze-validity rules, preprocessing code, preregistration, model code, synthetic data, and a de-identified feature-level benchmark where raw traces cannot be shared. Exact headset, runtime, locomotion method, frame rate, and eye-tracker calibration procedure should be reported.

\subsection*{Code and data availability}

The ArXiv source package includes the simulation code, fixed random seed, replicate-level synthetic results, summary table, and run metadata in the \texttt{anc/} directory. The plotted simulation figure is included at the package root for compilation. No human-participant data are reported or distributed with this manuscript.

\section{Philosophical Interpretation}

QSDD makes a modest but consequential philosophical move. It rejects the picture in which an observer passively extracts a fixed preference from an unchanged internal state. In an immersive environment, observation is an action situated in a perception-action loop. The route a person approaches determines the next view; the map they open changes what relations become salient; the question asked may alter subsequent judgment; and commitment changes the space of available actions.

This does not entail that reality is created by attention. It entails that the \emph{decision object} is partly constructed through context-sensitive interaction. The quantum-like formalism is useful when the available judgments cannot be treated as simultaneous readings from one stable joint distribution without loss of predictive structure.

The term ``superposition'' should likewise be read carefully. It does not mean that a person secretly chooses every route in parallel. It means that, relative to a specified measurement basis, the model represents a state that is not yet an eigenstate of the eventual decision. ``Collapse'' does not mean that consciousness changes a physical world. It denotes the conditional update associated with a registered outcome. ``Interference'' does not mean waves in neural tissue. It denotes a departure from classical additive decomposition under a constrained probabilistic model.

These translations preserve the intellectual force of the original philosophical fascination while relocating it to a research program that can fail. If cue order does not matter, if a classical latent-state model predicts better, or if the quantum-like parameters do not generalize, QSDD should be rejected or restricted.

\section{Limitations and Failure Criteria}

First, Hilbert-space models are expressive and can overfit. Geometric elegance is not evidence. The fixed-axis model reduces this risk, but richer QSDD variants will require explicit gauge constraints, recovery analysis, held-out prediction, and stronger classical baselines.

Second, behavioral equivalence is common. A quantum-like model and a classical model with memory can produce similar distributions. Noncommutativity is not exclusive to quantum models. The practical goal is not to prove a uniquely quantum ontology, but to determine whether the Hilbert-space model provides a more compact and general account under specified constraints.

Third, the simulation is conditional on its generating parameters, fixed initial azimuth, fixed cue-strength scale, independent Bernoulli observations, and simplified population structure. It demonstrates signed-cue observability and executability and exposes weak lapse identification; it does not establish real-world power or empirical support. A pilot-informed hierarchical simulation must precede the human study.

Fourth, telemetry is not transparent access to thought. Gaze, hesitation, and backtracking are multiply determined. Latent-state interpretations must survive manipulation checks and alternative emission models.

Fifth, spatial decisions can be continuous and high-dimensional. Discretizing them into left/right alternatives may discard geometry. Extensions should use higher-dimensional spaces, continuous POVMs, or hybrid state-space models while preserving identifiability.

Sixth, VR is a context in its own right. Results may not generalize to physical navigation, desktop interfaces, or expert scientific work. Cross-medium replication is therefore part of the research program, not an afterthought.

Finally, nothing in a successful QSDD experiment would establish consciousness-caused physical wave-function collapse. The following outcomes would count against the proposed framework: no replicable sequence effects under controlled exposure; failure of interference parameters to predict held-out conditions; poor parameter recovery; instability across equivalent task encodings; or equal/better generalization by simpler classical models.

\section{Future Work}

The empirical program proceeds through four dependent tests:
\begin{enumerate}[leftmargin=1.6em]
  \item \textbf{Pilot identification.} Estimate realistic cue strengths, participant heterogeneity, lapse, trial dependence, and tracking loss; rerun the design simulation before recruitment.
  \item \textbf{Controlled immersive evidence.} Test frozen cue-order predictions, protocol-qualified interference, progressive commitment, and reopening in a preregistered spatial task.
  \item \textbf{Telemetry-informed replication.} Integrate gaze, pose, and path emissions only after the choice model is identified; compare QSDD against classical sequential models on held-out participants and environments.
  \item \textbf{Quantum-domain transfer.} Evaluate circuit-inspection and debugging tasks with quantum learners and practitioners while preserving the same operator and measurement constraints.
\end{enumerate}

Each stage is independently falsifiable. Failure to recover parameters halts interpretation; failure against classical baselines restricts or rejects QSDD; and failure to transfer into circuit tasks limits the human--quantum interaction claim without invalidating a navigation-specific result.

Future work should also examine adaptive interfaces. If a model detects sustained unresolved competition, a system could offer a cue or alternative representation. Such intervention creates a closed loop and must be evaluated carefully: an interface that predicts uncertainty also changes it. Randomized micro-interventions can estimate this causal effect while avoiding the circular claim that a model is accurate because the interface forces behavior to match it.

\section{Conclusion}

QSDD converts a broad intuition about observation, context, and commitment into a model that can fail. Cue exposures are channels, registered judgments are instruments, route choice is an operational measurement, and a fixed sign-sensitive five-parameter geometry must predict new environments. The two-ground-truth simulation establishes that this repaired model is executable and distinguishable from a flexible classical comparator under stated conditions, while also revealing that lapse is weakly identified and that reliable discrimination may require more observations than the symmetry-restricted formulation suggested. It does not substitute for human evidence.

The decisive test is therefore empirical and comparative. A preregistered immersive study must recover the model, predict held-out order-contrast surfaces, survive classical latent-state and sequential-sampling alternatives, and respect the protocol assumptions required for interference-like claims. If those tests fail, QSDD should be restricted or rejected.

Human--quantum computer interaction supplies a consequential transfer domain. People must learn quantum concepts, inspect circuit representations, debug programs, interpret probabilistic outputs, and decide through imperfect tools. Methods from immersive interaction, visualization, behavioral telemetry, and empirical HCI can address that human layer directly. The resulting contribution is neither a theory of quantum consciousness nor a claim of hardware novelty; it is a falsifiable account of how representation order and embodied observation may shape decisions within spatial and quantum information environments.

\appendix
\section{Rotation Matrices and Implementation Details}

The Bloch-space rotations used in \cref{eq:fixed-rotations} are
\begin{equation}
R_x(a)=
\begin{pmatrix}
1&0&0\\0&\cos a&-\sin a\\0&\sin a&\cos a
\end{pmatrix},\qquad
R_y(a)=
\begin{pmatrix}
\cos a&0&\sin a\\0&1&0\\-\sin a&0&\cos a
\end{pmatrix}.
\end{equation}
They correspond to unitary qubit rotations $U_c=\exp[-i\omega_c\bm n_c\cdot\bm\sigma/2]$ under the standard Bloch representation. The route measurement uses $\Pi_L=\ket{L}\bra{L}$ and $\Pi_R=\ket{R}\bra{R}$. The dephasing multiplier $\kappa$ in the implementation is related to the mixture form
\begin{equation}
\Context_{\lambda}(\rho)=(1-\lambda)\rho+\lambda
\left(\Pi_L\rho\Pi_L+\Pi_R\rho\Pi_R\right)
\end{equation}
by $\kappa=1-\lambda$.

The implementation fixes $\phi_{\!*}=\pi/4$ in \cref{eq:bloch-initial}. This value is not optimized and therefore does not increase model dimension. Its nonzero sine and cosine break the reflection symmetries that would otherwise make one signed cue observationally even. A confirmatory study must freeze this constant before data collection rather than select it for fit.

For simulation, parameters were constrained to
\begin{equation}
\theta_0\in[0.15,2.99],\quad
\omega_L,\omega_M\in[0.05,2.50],\quad
\kappa\in[0.40,1],\quad
\epsilon\in[0,0.15].
\end{equation}
The classical coefficients were each constrained to $[-5,5]$. Optimization used L-BFGS-B with five QSDD and three classical starting points per dataset. These settings, the fixed azimuth, and all tested sample sizes are fully recorded in \texttt{anc/qsdd\_simulation.py}; they should be treated as analysis choices and subjected to sensitivity checks in an empirical study.

\section{Preregistration Skeleton}

\begin{table}[H]
\centering
\caption{Items to freeze before confirmatory data collection.}
\begin{tabularx}{\linewidth}{>{\bfseries}p{0.25\linewidth} X}
\toprule
Item & Required specification \\
\midrule
Primary contrast & Route-choice difference between forced $L\rightarrow M$ and $M\rightarrow L$ conflict trials \\
Commitment rule & Boundary geometry, dwell time $\tau$, and treatment of reversals \\
Trial exclusions & Tracking loss, calibration failure, premature movement, simulator withdrawal \\
Feature extraction & Gaze validity, ROI intersection, pause and backtrack thresholds, smoothing \\
Model parameterization & State dimension, operator constraints, priors, participant hierarchy, lapse model \\
Classical baselines & Logistic/Bayesian, hidden-state/POMDP, and drift-diffusion specifications \\
Validation & Simulation recovery, held-out split, ELPD/calibration metrics, ablations \\
Sample size & Simulation-based design analysis and stopping rule \\
Multiplicity & Confirmatory outcomes and correction or hierarchical strategy \\
Open materials & Scene, event schema, analysis code, synthetic data, and permitted de-identified outputs \\
\bottomrule
\end{tabularx}
\end{table}

\bibliographystyle{plainnat}
\bibliography{quantum_spatial_decision_dynamics}

\end{document}